\documentclass[preprint]{aastex}

\newcommand{\vecF}{\mbox{\boldmath $F$}}
\newcommand{\vecL}{\mbox{\boldmath $L$}}
\newcommand{\vecN}{\mbox{\boldmath $N$}}
\newcommand{\vecT}{\mbox{\boldmath $T$}}
\newcommand{\vecs}{\mbox{\boldmath $s$}}
\newcommand{\vecr}{\mbox{\boldmath $r$}}
\newcommand{\vecdels}{\mbox{\boldmath $\Delta s$}}
\newcommand{\vecrhat}{\mbox{\boldmath $\hat r$}}
\newcommand{\vecxhat}{\mbox{\boldmath $\hat x$}}
\newcommand{\vecyhat}{\mbox{\boldmath $\hat y$}}
\newcommand{\veczhat}{\mbox{\boldmath $\hat z$}}
\newcommand{\au}{\,{\rm AU}}
\newcommand{\yr}{\,{\rm yr}}
\newcommand{\Myr}{\,{\rm Myr}}

\begin{document}

\title{Evolution of the Obliquities of the Giant Planets in Encounters
       during Migration}

\author{Man Hoi Lee $^{\rm a}$, S. J. Peale $^{\rm a}$,
        Eric Pfahl $^{\rm b}$, William R. Ward $^{\rm c}$}
\affil{$^{\rm a}$ Department of Physics, University of California,
                  Santa Barbara, CA 93106}
\affil{$^{\rm b}$ Kavli Institute for Theoretical Physics,
                  University of California, Santa Barbara, CA 93106}
\affil{$^{\rm c}$ Space Studies Department,
                  Southwest Research Institute, Boulder, CO 80302}

\begin{abstract}
Tsiganis et al. [Tsiganis, K., et al., 2005. Nature 435, 459--461]
have proposed that the current orbital architecture of the outer solar
system could have been established if it was initially compact and
Jupiter and Saturn crossed the 2:1 orbital resonance by divergent
migration.
The crossing led to close encounters among the giant planets, but the
orbital eccentricities and inclinations were damped to their current
values by interactions with planetesimals.
Brunini [Brunini, A., 2006. Nature 440, 1163--1165] has presented
widely publicized numerical results showing that the close encounters
led to the current obliquities of the giant planets.
We present a simple analytic argument which shows that the change in
the spin direction of a planet relative to an inertial frame during an
encounter between the planets is very small and that the change in the
obliquity (which is measured from the orbit normal) is due to the
change in the orbital inclination.
Since the inclinations are damped by planetesimal interactions on
timescales much shorter than the timescales on which the spins precess
due to the torques from the Sun, especially for Uranus and Neptune,
the obliquities should return to small values if they are small before
the encounters.
We have performed simulations using the symplectic integrator SyMBA,
modified to include spin evolution due to the torques from the Sun and
mutual planetary interactions.
Our numerical results are consistent with the analytic argument for no
significant remnant obliquities.
\end{abstract}

\section{INTRODUCTION}

One of the fundamental questions of solar system formation is the
origin of planetary spins.
The spin axes of three of the four giant planets in our solar system
are tilted significantly, with the present values of the obliquities
(the angle between the spin axis and the orbit normal) of Jupiter,
Saturn, Uranus, and Neptune equal to $3\fdg1$, $27^\circ$, $98^\circ$,
and $30^\circ$, respectively.
\citet{bru06a} has recently proposed a unified mechanism for the origin
of the obliquities of the giant planets.
The starting point of Brunini's model is the so-called Nice model
\citep{tsi05} for the establishment of the orbital architecture of the
giant planets.
In the Nice model, the outer solar system was initially compact, with
Jupiter and Saturn closer than their mutual 2:1 mean-motion resonance
and all of the giant planets within $\sim 18\au$ of the Sun.
The scattering of planetesimals initially in a disk starting just
beyond the orbits of the planets caused Jupiter to migrate inward and
Saturn, Uranus, and Neptune outward.
When Jupiter and Saturn crossed the 2:1 mean-motion resonance, their
orbital eccentricities were excited and the orbits of Uranus and
Neptune were destabilized.
A phase of close encounters among the giant planets followed, and
Uranus and Neptune were scattered outward.
The orbital eccentricities and inclinations of the giant planets were
eventually damped to their present values by the interactions with
planetesimals.
\citet{bru06a} has performed numerical simulations of the Nice model
that follow the evolution of the planetary spin axes.
He found that large obliquities are generated during the encounter
phase and that the final obliquities are similar to the present
values, including the $98^\circ$ obliquity of Uranus.

\citet{bru06b} has subsequently retracted his paper on the basis that
simulations performed with another numerical code do not reproduce the
previous results and show changes in the obliquities of only a few
degrees in most runs.
However, he did not provide any physical explanation as to why the
later numerical results are more likely to be correct.
Given the wide publicity of the first results, we thought it important
to show in this paper a simple analytic argument (Section 2) that
explains why significant obliquities of the outer planets cannot
result from close encounters between these planets in the Nice model.
The analytic argument is verified by numerical simulations (Section
3).
We note that a summary of our results was presented by \citet{lee06}
prior to the retraction by \citet{bru06b}.
While our focus is on the Nice model, our analysis of the obliquity
changes in encounters can also be adapted to any scenario involving
close encounters among the planets, such as the scenario proposed by
\citet[][see also \citealt{chi06}]{gol04}, in which several ice giants
were formed in the outer solar system and all but two were ejected by
encounters.

\section{ANALYTIC ARGUMENT}

Let us consider the change in the spin direction of a planet of mass
$M$, radius $R$, and spin frequency $\omega$, due to an encounter with
a perturbing planet of mass $M_{\rm pert}$.
If we assume principle axis rotation, the spin angular momentum of the
planet is
\begin{equation}
\vecL = (\lambda + \ell) M R^2 \omega \vecs ,
\end{equation}
and the torque from the perturber at a distance $r$ in the direction
$\vecrhat$ is
\begin{equation}
\vecN = 3 {G M_{\rm pert}\over r^3} (J_2 + q') M R^2
        (\vecrhat \cdot \vecs) (\vecrhat \times \vecs) ,
\end{equation}
where $\vecs$ is the unit vector of the spin direction, $\lambda$ is
the moment of inertia of the planet normalized to $M R^2$, and $J_2$
is the quadrupole coefficient of the planet.
The torque from the planet's oblate figure on a satellite locks the
satellite to the planet's equator plane if the nodal precession period
of the satellite is short compared to the timescale on which the spin
direction of the planet is changed \citep{gol65}.
In the above equations, $\ell$ is the orbital angular momentum
(normalized to $M R^2 \omega$) of the satellites whose orbits are
coupled to the planet's equator, and $q'/J_2$ is the ratio of the
torque from the perturber on the satellites to that directly exerted
on the planet (see \citealt{war04} for explicit definitions of $\ell$
and $q'$).
From $d\vecL/dt = \vecN$, the rate of change of the spin direction is
\begin{equation}
{d\vecs \over dt} = 2 \alpha \left(M_{\rm pert} \over M_0\right)
                    \left( A \over r\right)^3
                    (\vecrhat \cdot \vecs) (\vecrhat \times \vecs) ,
\label{dsdt}
\end{equation}
where $M_0$ is the mass of the Sun and
\begin{equation}
\alpha = {3 G M_0 \over 2 \omega A^3}
         \left(J_2 + q' \over \lambda + \ell\right)
\end{equation}
is the precession constant of the planet (plus its satellites) at the
{\it current} orbital semimajor axis, $A$, of the planet.

Because the torque decreases rapidly with the distance $r$, we expect
the total change in the spin direction in an encounter, $|\vecdels|$,
to be roughly equal to the rate of change at closest approach,
$|d{\vecs}/dt|_{r_p}$, times the duration of closest approach,
$r_p/v_p$:
\begin{equation}
|\vecdels| \sim \left|{d{\vecs}\over dt}\right|_{r_p}
                        {r_p \over v_p}
           \propto \alpha \left(M_{\rm pert} \over M_0\right)
                   \left(A \over r_p\right)^2 {A \over v_p} ,
\end{equation}
where $r_p$ and $v_p$ are, respectively, the relative separation and
speed at closest approach and the proportionality constant is a
function of the encounter geometry.
A more accurate expression for $\vecdels$ can be obtained by
neglecting solar gravity and using the two-body approximation during
the encounter.
Then
\begin{equation}
\vecdels = \int_{-\infty}^{+\infty} {d\vecs \over dt} dt
         = \int_{-\arccos(-1/e)}^{+\arccos(-1/e)} {d\vecs \over dt}
           \left(df \over dt\right)^{-1} df ,
\label{deltas}
\end{equation}
where $e$($\ge 1$) and $f$ are, respectively, the eccentricity and
true anomaly of the relative orbit.
In the coordinate system where the relative orbit is in the $x$-$y$
plane and the closest approach is in the $x$ direction,
the integration in Eq. (\ref{deltas}) yields
\begin{eqnarray}
\vecdels &=& \alpha \left(M_{\rm pert} \over M_0\right)
             \left(A \over r_p\right)^2 {A \over v_p} {1 \over 1+e}
\nonumber \\
         & & \times \Bigg\{
             \left[\arccos(-1/e) + {e \over 3} \left(2 + 1/e^2\right)
             \left(1 - 1/e^2\right)^{1/2}\right] \sin 2\theta
             \sin\phi \vecxhat
\nonumber \\
         & & \qquad
            -\left[\arccos(-1/e) + {e \over 3} \left(4 - 1/e^2\right)
             \left(1 - 1/e^2\right)^{1/2}\right] \sin 2\theta
             \cos\phi \vecyhat
\nonumber \\
         & & \qquad
            +{2 e \over 3} \left(1 - 1/e^2\right)^{3/2} \sin^2\theta
             \sin2\phi \veczhat
             \Bigg\} ,
\end{eqnarray}
where $\theta$ and $\phi$ are the spherical coordinate angles
specifying the orientation of the spin axis in the $xyz$ system:
$\vecs = \sin\theta \cos\phi \vecxhat + \sin\theta \sin\phi \vecyhat +
\cos\theta \veczhat$.
Contour plots of the magnitude, $|\vecdels|$, as a function of
$\theta$ and $\phi$ show that $|\vecdels|$ reaches equal maximum
values at four points with $\theta = 45^\circ$ or $135^\circ$ and
$\phi = 0^\circ$ or $180^\circ$ for any $e > 1$.
For $e = 1$ (i.e., parabolic encounter), $|\vecdels| = (\pi/2) \alpha
(M_{\rm pert}/M_0) (A/r_p)^2 (A/v_p) |\sin 2\theta|$, which is
independent of $\phi$ and maximum at $\theta = 45^\circ$ or
$135^\circ$.
The plot of the maximum $|\vecdels|$ for a given $e$ as a function of
$e$ shows that it decreases monotonically with increasing $e$.
Thus $|\vecdels|$, as a function of $e$, $\theta$, and $\phi$, is
maximum for $e = 1$ and $\theta = 45^\circ$ or $135^\circ$:
\begin{equation}
|\vecdels|_{\rm max} = {\pi \over 2} \alpha
                       \left(M_{\rm pert} \over M_0\right)
                       \left(A \over r_p\right)^2 {A \over v_p} .
\label{deltasmax}
\end{equation}

Since Uranus has the largest current obliquity and the largest changes
in its spin direction come from encounters with Saturn, we consider
first the change in the spin direction of Uranus due to an encounter
with Saturn.
As we shall see in Figs. \ref{fig1} and \ref{fig2}, Saturn's orbit (of
semimajor axis $a_{\rm S}$) remains nearly circular, and an encounter
between Saturn and Uranus must have Uranus on an eccentric orbit
(of eccentricity $e_{\rm U}$), with Uranus usually near its
perihelion.
Thus the relative encounter speed at large separation is
$v_0 \approx (G M_0/a_{\rm S})^{1/2} [(1+e_{\rm U})^{1/2} - 1]
\la 0.22 (G M_0/a_{\rm S})^{1/2}$ if $e_{\rm U} \la 0.5$.
Since the encounter phase occurs long after the formation of the
regular satellites in the Nice model, the regular satellite systems
must survive the encounter, and a reasonable lower limit to the
separation $r_p$ at closest approach is the semimajor axis
$a_{\rm Titan}$ of the orbit of Titan around Saturn.
For $r_p = a_{\rm Titan}$, $2 G (M_{\rm S} + M_{\rm U})/r_p \gg v_0^2$
(where $M_{\rm S}$ and $M_{\rm U}$ are the masses of Saturn and
Uranus, respectively), and
$v_p = [v_0^2 + 2 G (M_{\rm S} + M_{\rm U})/r_p]^{1/2}
\approx [2 G (M_{\rm S} + M_{\rm U})/r_p]^{1/2}$.
So the maximum change in the spin direction of Uranus from an
encounter with Saturn is
\begin{equation}
|\vecdels|^{\rm S}_{\rm max,U} = 0\fdg33 (a_{\rm Titan}/r_p)^{3/2} ,
\end{equation}
if we adopt $\alpha_{\rm U} = (4.6\Myr)^{-1}$ from \citet{tre91}.

For the same encounter between Uranus and Saturn, the maximum change
in the spin direction of Saturn is $|\vecdels|^{\rm U}_{\rm max,S} =
0\fdg11 (a_{\rm Titan}/r_p)^{3/2}$ if $\alpha_{\rm S} =
(0.26\Myr)^{-1}$ \citep{tre91}.
If we also adopt $\alpha_{\rm N} = (17.6\Myr)^{-1}$ from
\citet{tre91}, the changes from an encounter between Neptune and
Saturn are nearly identical to those from an encounter between Uranus
and Saturn, with
$|\vecdels|^{\rm S}_{\rm max,N} \approx |\vecdels|^{\rm S}_{\rm max,U}$
and
$|\vecdels|^{\rm N}_{\rm max,S} \approx |\vecdels|^{\rm U}_{\rm max,S}$,
because Uranus and Neptune have similar masses and
$\alpha_{\rm U} A_{\rm U}^3 \approx \alpha_{\rm N} A_{\rm N}^3$ (see
Eq. (\ref{deltasmax})).
For an encounter between Uranus and Neptune, 
$|\vecdels|^{\rm N}_{\rm max,U} \approx |\vecdels|^{\rm U}_{\rm max,N}
\approx 0\fdg3 (a_{\rm Oberon}/r_p)^{3/2}$.
We do not consider encounters with Jupiter, since such encounters
usually result in systems that do not resemble the solar system by,
e.g., ejecting the planet encountering Jupiter \citep{tsi05}.

The precession constants adopted above include the contributions from
the satellites of the planets.
In the case of Uranus, the contributions are mainly from Titania and
Oberon.
However, the nodal precession periods of Titania and Oberon ($\ga
1300\yr$ due to the oblateness of Uranus alone and $\sim 200\yr$ with
the inclusion of the secular interactions among the satellites;
\citealt{las87}) are much longer than the encounter timescale with
either Saturn ($r_p/v_p \sim 2\,$days for $r_p \sim a_{\rm Titan}$) or
Neptune ($r_p/v_p \sim 1\,$day for $r_p \sim a_{\rm Oberon}$), and
their orbits would not be able to follow the equator of Uranus during
the encounter \citep{gol65}.
This would reduce $\alpha_{\rm U}$ and hence the already small maximum
change in the spin direction of Uranus by a factor of $5.3$.
Similarly, without the contributions from the satellites to the
precession constants, the maximum changes in the spin directions of
Neptune and Saturn are reduced by a factor of $5.7$ and $3.7$,
respectively.
Then the maximum change in the spin direction is $\la 1^\circ$ for any
pair-wise encounter among Saturn, Uranus, and Neptune, even if we
consider an extremely close encounter with $r_p$ equal to twice the
sum of the planetary radii, which the regular satellite systems are
unlikely to survive.

We have just shown that the change in the spin direction in
{\it inertial space} during an encounter between the planets is very
small.
Thus any change in the obliquity (which is defined as the angle
between the spin axis and the orbit normal) must be due to a change in
the orbital inclination.
As we shall see in, e.g., Figs. \ref{fig1} and \ref{fig2}, the orbital
inclinations of the ice giants can reach maximum values of $\sim
10^\circ$--$15^\circ$ during the encounter phase.
However, because the timescale on which the inclination is changed by
encounters and damped by planetesimals ($\la 1\Myr$) is much less than
the precession period of the spin axis about the orbit normal due to
solar torque ($\ga 2\pi/\alpha_{\rm U} = 29\Myr$ for Uranus),
there is not enough time for the spin axis to precess much due to
solar torque, and the spin axis does not stray far from its initial
position nearly perpendicular to the initial orbital planes of the
planets.
Therefore, we expect the obliquities to return to small values as the
inclinations are damped to their present values.

\section{NUMERICAL SIMULATIONS}

To verify the analytic argument in Section 2 and to eliminate the
possibility of unexpected secular spin-orbit effects, we have
performed numerical simulations of the Sun and the four giant planets
using the symplectic $N$-body code SyMBA \citep{dun98}, which we have
modified to include spin evolution.
For the orbital evolution, we ignore the negligible effects of the
oblateness of the planets, and we use imposed migration and damping of
eccentricity and inclination to model the effects of the
planetesimals.
\citet{lee02} have described how SyMBA can be modified to include
forced migration and damping of eccentricity and inclination.
The unit vector $\vecs_k$ of the spin direction of planet $k$ is
evolved according to
\begin{eqnarray}
{d\vecs_k \over dt} &=& 2 \alpha_k \left(A_k \over r_{0k}\right)^3
                        (\vecrhat_{0k} \cdot \vecs_k)
                        (\vecrhat_{0k} \times \vecs_k)
\nonumber \\
                    & & + 2 \alpha_k \sum_{j>0, j \neq k}
                        \left(M_j \over M_0\right)
                        \left(A_k \over r_{jk}\right)^3
                        (\vecrhat_{jk} \cdot \vecs_k)
                        (\vecrhat_{jk} \times \vecs_k) ,
\label{dskdt}
\end{eqnarray}
where the symbols are as in Eq. (\ref{dsdt}), but with the subscript 0
for the Sun and $j,k \neq 0$ for the planets.
In Eq. (\ref{dskdt}), the first term is due to the torque from the
Sun and the second term is due to the torques from the other planets.
The recursively subdivided time step used by SyMBA to handle close
encounters between the planets is also implemented for the spin
evolution due to the planetary torques.
In the Appendix, we describe in more details how SyMBA is modified to
include spin evolution and the tests that were performed.

\begin{figure}
\epsscale{1.0}
\plotone{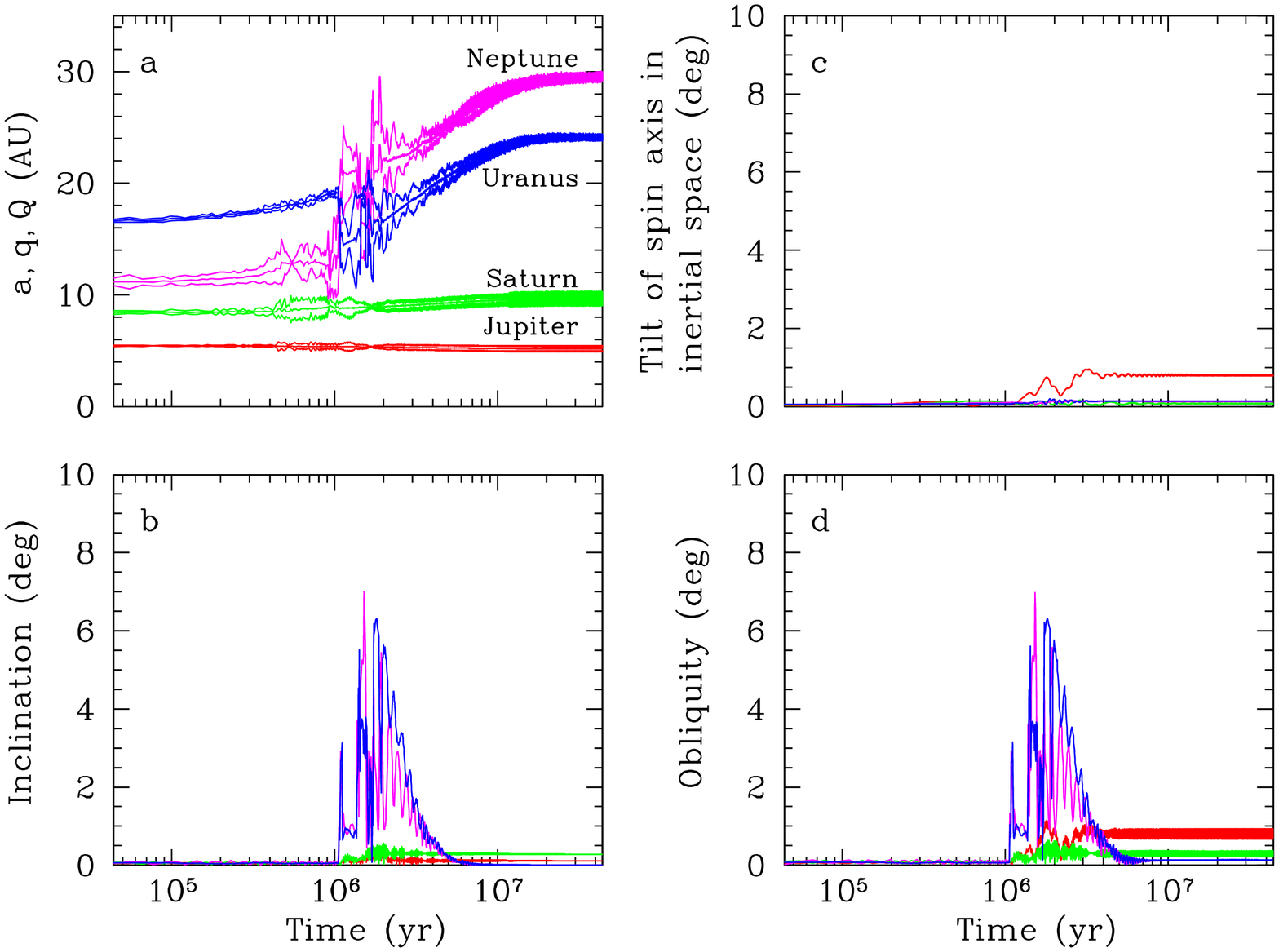}
\caption{
Evolution of the orbits and spin axes of the giant planets in a
simulation from series I (see text for the details of the setup).
(a) Orbital semimajor axis ($a$) and the minimum ($q$) and
maximum ($Q$) heliocentric distances.
Note that the ice giants switch positions in this simulation.
(b) Orbital inclination.
(c) Tilt of the spin axis in inertial space.
(d) Obliquity.
The planets are Jupiter (red), Saturn (green), Uranus (blue), and
Neptune (magenta).
The orbital inclination and the tilt of the spin axis in inertial
space are measured relative to the $z$-axis of the inertial frame,
which is nearly perpendicular to the initial orbital planes of the
planets, while the obliquity is the angle between the spin axis and
the orbit normal.
The directions of the spin axes show very little change in inertial
space.
The obliquities reflect primarily the inclinations and return to
small values as the inclinations are damped.
\label{fig1}}
\end{figure}

\begin{figure}
\epsscale{1.0}
\plotone{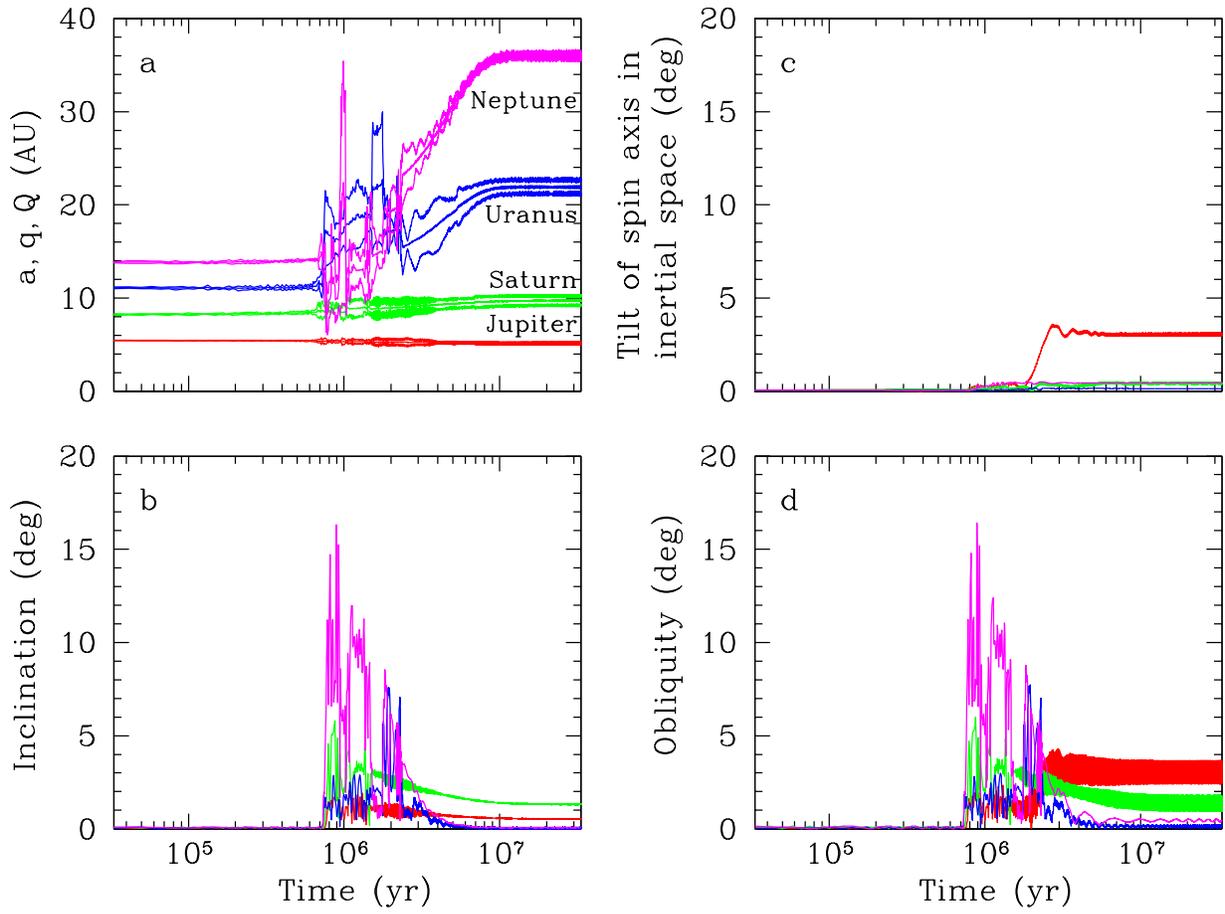}
\caption{
Same as Fig. \ref{fig1}, but for a simulation from series III.
The ice giants do not switch positions in this simulation.
\label{fig2}}
\end{figure}

We present the results from 4 series of runs, each with 30
simulations.
The simulations were performed with a time step of $0.125\yr$.
For the precession constants at the current orbital semimajor axes, we
adopt $\alpha_{\rm J} = (0.077\Myr)^{-1}$ for Jupiter ,
$\alpha_{\rm S} = (0.26\Myr)^{-1}$ for Saturn, $\alpha_{{\rm I}_1} =
\alpha_{\rm U} = (4.6\Myr)^{-1}$ for the initially inner ice giant,
and $\alpha_{{\rm I}_2} = \alpha_{\rm N} = (17.6\Myr)^{-1}$ for the
initially outer ice giant.
These values include the contributions from the satellites
\citep{tre91}.
Although the inner and outer ice giants can switch positions (see,
e.g., Fig. \ref{fig1}), the fact that we adopt the parameters of
Uranus (Neptune) for the initially inner (outer) ice giant is not
important, because Uranus and Neptune have similar masses and
$\alpha_{\rm U} A_{\rm U}^3 \approx \alpha_{\rm N} A_{\rm N}^3$.
The four series of runs have the following initial conditions in
common: the orbital eccentricities are $10^{-3}$; both the orbit
normals and the spin axes are tilted by $10^{-3}$ radians (or
$0\fdg057$) from the $z$-axis of the inertial frame; the mean
anomalies, the arguments of perihelion, and the longitudes of the
ascending nodes on the $x$-$y$ plane of both the orbits and the
equators are random variables.

In series I, the initial orbital semimajor axes and the imposed
migration and damping of eccentricity and inclination are similar to
those adopted by \citet{bru06a}.
The initial semimajor axis of Jupiter is $a_{\rm J} = 5.45\au$.
The initial semimajor axis of Saturn is varied in the range
$8.38$--$8.48\au$, whereas those of the ice giants are varied in the
ranges $9.9$--$12\au$ and $13.4$--$17.1\au$, while keeping the initial
orbital separation of the ice giants larger than $2\au$.
The imposed migration rate is
\begin{equation}
{\dot a}_k = {\Delta a_k \over \tau} \exp(-t/\tau) ,
\label{dakdt}
\end{equation}
where $\Delta a_k$ is the difference between the current semimajor
axis, $A_k$, and the initial semimajor axis.
The migration timescale $\tau$ is varied between $1$ and $10\Myr$, and
the total time span of a simulation is $10\tau$.
The imposed eccentricity and inclination damping rates are
\begin{equation}
{\dot e}_k = -e_k/(2 \tau_e), \quad {\dot i}_k = -i_k/(2 \tau_i) ,
\end{equation}
respectively, with $\tau_e = \tau_i = \tau/10$.
The eccentricity and inclination dampings are applied only if the
orbital distance of a planet is between just outside the initial orbit
of the outer ice giant and $30\au$.
Series II is similar to series I, but with $\tau_e = \tau_i =
\tau/20$.

Series III and IV have initial orbital semimajor axes closer to those
adopted by \citet{tsi05} and use a different prescription for
migration and damping.
The initial $a_{\rm J} = 5.45\au$, and the initial semimajor axes of
Saturn and the inner and outer ice giants are varied in the ranges
$8.0$--$8.5\au$, $11$--$13\au$, and $13.5$--$17\au$, respectively,
with the ice giants at least $2\au$ apart.
The main difference from series I and II is that the semimajor axis of
the inner ice giant is about $1\au$ larger.
For the imposed migration, we use Eq. (\ref{dakdt}) for Jupiter and
Saturn and
\begin{equation}
{\dot a}_k = {\Delta a_k \over \tau} \left(t \over \tau\right)
             \exp[-t^2/(2\tau^2)]
\label{dakdtice}
\end{equation}
for the ice giants.
With the inner ice giant further from Saturn initially, the migration
rate in Eq. (\ref{dakdtice}) ensures that the ice giants do not
migrate too far by the time of the Jupiter-Saturn 2:1 resonance
crossing, as seen in the simulations of \citet{tsi05}.
For all planets, the damping rates are
\begin{equation}
{\dot e}_k/e_k = -K_e |{\dot a}_k/a_k| , \quad
{\dot i}_k/i_k = -K_i |{\dot a}_k/a_k| ,
\end{equation}
and they are applied at all times.
This form of damping has been used to model planet-disk interactions
(e.g., \citealt{lee02}), and it has the reasonable property of tying
the damping rates to the migration rate (i.e., less damping when the
migration slows down).
We adopt $K_e = K_i = 10$ for series III and $K_e = K_i = 5$ for
series IV.

The fractions of cases with all four giant planets surviving on
reasonable orbits are $47\%$, $73\%$, $57\%$, and $50\%$ in series
I--IV, respectively.
They are all roughly consistent with the $67\%$ reported by
\citet{tsi05}, given the small number of simulations in Tsiganis et
al. ($43$ simulations) and in each of our series ($30$ simulations).
The results from two typical simulations with all four giant planets
surviving on reasonable orbits --- one from series I and the other
from series III --- are shown in Figs. \ref{fig1} and \ref{fig2}.
Fig. \ref{fig1} is an example where the ice giants switch positions.
In each Figure, panel a shows the evolution of the orbital semimajor
axes $a$, the perihelion distances $q = a(1-e)$, and the aphelion
distances $Q = a(1+e)$, while panels b, c, and d show the evolution of
the orbital inclinations $i$, the tilts of the spin axes from the
$z$-axis of the inertial frame (which is nearly perpendicular to the
initial orbital planes of the planets), and the obliquities
$\varepsilon$, respectively.
It can be seen from the correlation of the changes in panels b and d
of each Figure that large changes in the obliquities of Saturn and
the ice giants during close encounters are due to changes in orbital
inclinations.
The changes in the directions of the spin axes relative to inertial
space during the encounters are minuscule (panel c).
As the inclinations are damped to small values, the spin axes keep
their orientations in inertial space, and the obliquities are
similarly decreased.
These behaviors are exactly what we expect from the analytic argument
in Section 2.

For Jupiter, we note that both the obliquity and the tilt of the spin
axis from the inertial $z$-axis increase to $1$--$3$ degrees in
Figs. \ref{fig1} and \ref{fig2}.
This could be due to secular spin-orbit coupling, because the spin
precession due to solar torque is quite fast for Jupiter
($\alpha_{\rm J} = (0.077\Myr)^{-1}$), but a detailed examination of
this process is beyond the scope of the present paper.

Among the cases with all four giant planets surviving on reasonable
orbits ($68$ out of $120$ simulations in all four series), there are
only four cases where one or both of the ice giants have final
obliquities greater than $5^\circ$, including one case where one of
the ice giants (Neptune) has a final obliquity greater than $10^\circ$
($\varepsilon_{\rm N} = 27^\circ$).
However, these cases involve one or two extremely close encounters
between the planets (usually between Saturn and the initially inner
ice giant), for which the regular satellite systems are unlikely to
survive.
Even if the regular satellites could survive these encounters, their
orbits would not be able to follow the equators of the planets during
the encounters, and the final obliquities obtained in our simulations
(which include the contributions from the satellites) should be
reduced by a factor of $5$--$6$ (see Section 2) and $< 5^\circ$ even
in the most extreme case.

\section{CONCLUSIONS}

We have analyzed the evolution of the obliquities of the giant planets
in the Nice model \citep{tsi05}, where the Jupiter-Saturn 2:1
resonance crossing led to close encounters among the giant planets and
the orbital eccentricities and inclinations were eventually damped to
their present values by the interactions with planetesimals.
We presented a simple analytic argument which shows that the change in
the spin direction of a planet in inertial space during an encounter
between the planets is very small and that the change in the obliquity
(which is measured from the orbit normal) is due to the change in the
orbital inclination.
Since the inclinations are changed by encounters and damped by
planetesimal interactions on timescales much shorter than the
timescales on which the spins precess due to solar torques, especially
for Uranus and Neptune, the obliquities return to small values
if they are small before the encounters.
We have performed simulations using the symplectic integrator SyMBA
modified to include spin evolution.
The numerical results are consistent with the analytic argument
leading to no significant remnant obliquities.
The analytic argument provides a physical explanation as to why the
numerical results reported by \citet{bru06a}, which are not confirmed
by later simulations by him using another numerical code
\citep{bru06b} or the simulations in this paper, are indeed incorrect.

Giant impacts in the late stages of the formation of Uranus and
Neptune remain a plausible explanation for the obliquities of the ice
giants, but a giant impact origin for Saturn's obliquity is
problematic because of its large mass and spin angular momentum
\citep{lis91,don93}.
Alternatively, the obliquities of the giant planets could be generated
by any process that twists the total angular momentum vector of the
solar system on a timescale of $\sim 0.5\Myr$ \citep{tre91}.
However, the similarity between Saturn's spin-axis precession rate and
the nodal regression rate of Neptune's orbit, and the near match of
Saturn's $27^\circ$ obliquity to that of the Cassini state 2 of the
secular spin-orbit resonance, strongly argue for the capture of Saturn
into the Cassini state whose obliquity increases during the dispersal
of the planetesimal disk \citep{war04,ham04}.

\acknowledgments
We thank H.~F.~Levison and K.~Tsiganis for informative discussions.
This research was supported in part by NASA grants NNG05GK58G and
NNG06GF42G (M.H.L. and S.J.P.), NASA grant NNG04GJ11G (W.R.W.), and
NSF grant PHY 99-07949 (E.P.).


\appendix

\section{NUMERICAL METHODS}

In this appendix we describe how the symplectic integrator SyMBA
\citep{dun98} is modified to include the evolution of the spin axes
according to Eq. (\ref{dskdt}) and the tests that were performed.
Our algorithm differs from the Lie-Poisson integrator for rigid body
dynamics by \citet{tou94} in assuming principle axis rotation and
handling close encounters.

SyMBA is based on a variant of the \citet{wis91} method, with the
gravitational $N$-body Hamiltonian written in terms of positions
relative to the Sun and barycentric momenta, and employs a multiple
time step technique to handle close encounters.
The potential of the gravitational interaction between planets $j$ and
$k$ is of the form $-G m_j m_k /r_{jk}$, which we simplify as $-1/r$
in the following discussion.
In SyMBA, a set of cutoff radii $R_1 > R_2 > \cdots$ is chosen, with
$R_1$ a few times the mutual Hill radius, and the potential $-1/r$ is
decomposed into $\sum_{\ell=0}^\infty V_\ell$, or equivalently the
force $\vecF$ into
$\sum_{\ell=0}^\infty \vecF_\ell
= - \sum_{\ell=0}^\infty \partial V_\ell/\partial \vecr
= - \sum_{\ell=0}^\infty P_\ell(r) \vecr/r^3$.
The properties of the partition functions $P_\ell(r)$ are such that
(i) $P_\ell$ (except $P_0$) increases smoothly from zero at $r <
R_{\ell+2}$ to one at $r = R_{\ell+1}$ and then decreases smoothly to
zero at $r \ge R_\ell$ and
(ii) $P_0$ increases smoothly from zero at $r < R_2$ to one at $r \ge
R_1$.
The force $\vecF_\ell$ is to be applied with a time step $h_\ell$,
with $h_\ell/h_{\ell+1}$ an integer ($= 3$ in the standard
implementation).
A single step of overall time step $h_0$ of the SyMBA algorithm can be
represented by the following sequence of substeps:
\begin{equation}
E_{\rm Sun}\left(h_0 \over 2\right)
E_{\rm int,0}\left(h_0 \over 2\right)
E_{\rm recur}(h_0)
E_{\rm int,0}\left(h_0 \over 2\right)
E_{\rm Sun}\left(h_0 \over 2\right) .
\label{SyMBA}
\end{equation}
In the substep $E_{\rm Sun}(h_0/2)$, each planet takes a linear drift
in its heliocentric position (corresponding to a linear drift in the
position of the Sun from the barycenter) for time $h_0/2$.
In the substep $E_{\rm int,0}(h_0/2)$, each planet receives a kick to
its linear momentum due to the $\ell = 0$ level forces from the other
planets for time $h_0/2$.
In the recursive substep $E_{\rm recur}(h_0)$, a planet that is not
having close encounters (i.e., a planet whose separation from any
other planet is greater than $R_1$) evolves along a Kepler orbit for
time $h_0$, while a pair of planets having an encounter have their
time steps recursively subdivided to a maximum level of
$\ell_{\rm max}$ --- with the force at level $\ell$, $\vecF_\ell$,
applied with a time step $h_\ell$ --- if their separation $r >
R_{\ell_{\rm max}+1}$ during the overall step (see \citealt{dun98} for
details).

In Eq. (\ref{dskdt}), the torque on the spin axis of planet $k$ from
planet $j$ is of the form $2 \alpha_k (M_j/M_0) \times (A_k/r_{jk})^3
(\vecrhat_{jk} \cdot \vecs_k) (\vecrhat_{jk} \times \vecs_k)$, which
we simplify as $\vecT = r^{-3} (\vecrhat \cdot \vecs) (\vecrhat \times
\vecs)$ in this discussion.
As for the force $\vecF$, we decompose $\vecT$ into
$\sum_{\ell=0}^\infty \vecT_\ell
= \sum_{\ell=0}^\infty P_\ell(r) r^{-3} (\vecrhat \cdot \vecs)
 (\vecrhat \times \vecs)$
using the partition functions $P_\ell(r)$,
and $\vecT_\ell$ is to be applied with a time step $h_\ell$.
Thus our modified algorithm (with imposed migration and damping) can
be represented by the following sequence of substeps:
\begin{eqnarray}
E_{a}\left(h_0 \over 2\right)
E_{ei}\left(h_0 \over 2\right)
E_{\rm Sun}\left(h_0 \over 2\right)
E_{\rm Sun}^{\rm spin}\left(h_0 \over 2\right)
E'_{\rm int,0}\left(h_0 \over 2\right)
E'_{\rm recur}(h_0)
\qquad\qquad
\nonumber \\
E'_{\rm int,0}\left(h_0 \over 2\right)
E_{\rm Sun}^{\rm spin}\left(h_0 \over 2\right)
E_{\rm Sun}\left(h_0 \over 2\right)
E_{ei}\left(h_0 \over 2\right)
E_{a}\left(h_0 \over 2\right) .
\end{eqnarray}
In the substep $E_{a}(h_0/2)$, the orbital semimajor axis $a$ of each
planet is changed according to the exact solution to the imposed
migration rate (with all of the other orbital elements fixed) for time
$h_0/2$ (see \citealt{lee02} for details).
Similarly, the orbital eccentricity $e$ and inclination $i$ of each
planet are changed according to the exact solutions to the imposed
damping rates for time $h_0/2$ in the substep $E_{ei}(h_0/2)$.
In the substep $E_{\rm Sun}^{\rm spin}(h_0/2)$, the spin of each
planet is evolved due to the solar torque (i.e., the first term in
Eq. (\ref{dskdt})) alone for time $h_0/2$, which is a rotation of
$\vecs_k$ by an angle $2 \alpha_k (A_k/r_{0k})^3 (\vecrhat_{0k} \cdot
\vecs_k) h_0/2$ about $\vecr_{0k}$, because $\vecr_{0k}$ is fixed
during this substep.
The substep $E'_{\rm int,0}(h_0/2)$ is the same as
$E_{\rm int,0}(h_0/2)$ in the original SyMBA algorithm, but it also
evolves the spin direction of each planet due to the $\ell = 0$ level
torques from the other planets for time $h_0/2$.
Unlike the substep applying the solar torque alone, there is not an
exact solution for the change in the spin direction with the torques
from multiple planets, and we use the midpoint method.
Similarly, $E'_{\rm recur}(h_0)$ is the same as $E_{\rm recur}(h_0)$,
but it also applies the torque at level $\ell$, $\vecT_\ell$, with a
time step $h_\ell$, recursively down to the maximum level
$\ell_{\rm max}$, for a pair of planets having an encounter.

We have tested our implementation of the spin evolution with several
tests.
In the first set of tests, we checked that the code produces the
correct spin axis precession around the orbit normal in the presence
of the solar torque alone.
We integrated Jupiter alone with a variety of orbital eccentricity $e$
(up to 0.2).
We also integrated the four giant planets with their current $e$ and
$i$, but with their masses reduced by a factor of $10^{-10}$ (so that
they are on unperturbed Kepler orbits).
A variety of obliquities $\varepsilon$ were used, and we integrated
for $10^7\yr$ using a time step of $0.25\yr$.
We found a small variation in $\varepsilon$ (up to $\sim 10^{-3}$
degrees) over the period of an orbit, which is expected since
Eq. (\ref{dskdt}) is not orbit-averaged, but there was no detectable
secular change in $\varepsilon$.
The average spin precession rate (measured by the total change in the
longitude of the ascending node of the equator on the orbital plane)
agreed with the analytic result $\alpha (1-e^2)^{-3/2} \cos\varepsilon$
to better than one part in $5 \times 10^4$.

In the second set of tests, we checked that the code produces the
correct capture into (and escape from) secular spin-orbit resonance.
The simulations were simplified versions of the ones in \citet{ham04}.
We integrated Saturn alone with a time step of $0.58\yr$.
Initially, $a = 9.54\au$, $e = 10^{-3}$ and $i = 0\fdg064$ (the last
being the amplitude associated with the perturbation to Saturn's
orbital plane due to the nodal regression of Neptune).
We imposed an orbital nodal precession rate of $g = -\alpha (0.89 +
0.23 e^{-t/\tau})$ (magnitude decreasing with time) or $g = -\alpha
(1.12 - 0.23 e^{-t/\tau})$ (magnitude increasing with time), with
$\tau = 2.5$--$10 \times 10^8\yr$.
In the cases where $|g|$ was decreased from above $\alpha$ to below
$\alpha$, Saturn's spin was captured into Cassini state 2, with the
obliquity librating about the analytic result in \citet{war04}.
In the cases where $|g|$ was increased from below $\alpha$ to above
$\alpha$, the spin was first captured into Cassini state 1, with the
obliquity librating about the analytic result in \citet{war04}.
There was a jump in the obliquity when Cassini state 1 disappeared,
and the obliquity subsequently oscillated about a constant value.
The results agreed with the analytic theory and numerical results of
\citet{war04} and \citet{ham04}.

As a very stringent test of the spin evolution due to close mutual
planetary interactions, we performed simulations similar to the binary
planet test described in \citet{dun98}, where two planets are in a
bound orbit and their center of mass orbits a star.
The center of mass of the binary was placed in a $5.2\au$ circular
orbit about a solar-mass star.
The initial semimajor axis of the relative orbit of the binary was
$0.0125\au$ and the initial eccentricity was $0.4$.
One planet had a mass of $M_1 = 0.9 \times 10^{-3}M_\sun$ and the
other $M_2 = 1.1 \times 10^{-3}M_\sun$.
Both planets had the same precession constant as Jupiter, and their
initial obliquities were $\varepsilon_1 = 60^\circ$ and $\varepsilon_2
= 30^\circ$.
We integrated this system for $500\yr$ (or about 16000 orbital
periods of the binary).
Because the torques of the planets on each other dominated over the
torques from the Sun, the average precession rates of the spin axes
around the normal to the relative orbit are known analytically, if we
ignore the solar torques.
We found that the numerical precession rates agreed with the analytic
results to within 0.0008 and 0.0022 fractionally for planets 1 and 2,
respectively.
The small discrepancies were not sensitive to the time step (we used
$0.25\yr$ and smaller), and they are likely real and due to solar
perturbations.
Solar perturbations also caused the eccentricity of the relative orbit
to vary between $0.4$ and $0.3964$ and the obliquities to show small
variations on the same timescale.


\clearpage

\end{document}